\documentclass[12pt]{article}

\newcommand{\beq}[1]{\begin{equation}\label{#1}}
\newcommand{\eeq}{\end{equation}}
\newcommand{\bear}[1]{\begin{eqnarray}\label{#1}}
\newcommand{\ear}{\end{eqnarray}}

\renewcommand{\thesection}{\arabic{section}}
\renewcommand{\theequation}{\arabic{section}.\arabic{equation}}
\catcode`\@=11 \@addtoreset{equation}{section}\catcode`\@=12

\newcommand{\np}{ {\newpage } }
\newcommand{\R}{ \mbox{\rm I$\!$R} }

\newcommand{\e}{ \mbox{\rm e} }
\newcommand{\p}{\partial}

\newcommand{\btu}{\bigtriangleup}

\newcommand{\const}{\mathop{\rm const}\nolimits}
\newcommand{\fnt}{\footnotetext}
\newcommand{\fnm}{\footnotemark}
\newcommand{\Picture}[4]{{\small\begin{figure}[h]
\begin{center}\begin{picture}(#1,#2)
\put(0,#2){\special{em:graph #3}}
\end{picture}
\addtocounter{figure}{1}
\par\bigskip
\figurename~\thefigure. #4
\end{center}
\end{figure}}}

\begin{document}

\thispagestyle{empty}

\begin{center}{ \large \bf
Billiard representation for
multidimensional multi-scalar cosmological model
 with exponential potentials}

\end{center}

\bigskip

\begin{center}

\vspace{15pt}

\normalsize\bf

H. Dehnen \fnm[1]\fnt[1]{Heinz.Dehnen@uni-konstanz.de},

\it Universit$\ddot{a}$t Konstanz, Fakult$\ddot{a}$t f$\ddot{u}$r Physik,
Fach  M 568, D-78457, Konstanz \\

\vspace{5pt}

\bf V.D. Ivashchuk\fnm[2]\fnt[2]{ivas@rgs.phys.msu.su}
and  V.N. Melnikov \fnm[3]\fnt[3]{melnikov@rgs.phys.msu.su}

\it Center for Gravitation and Fundamental Metrology,
VNIIMS, 3/1 M. Ulyanovoy Str.,
Moscow 119313, Russia  and\\
Institute of Gravitation and Cosmology, PFUR,
6 Miklukho-Maklaya Str., \\ Moscow 117198, Russia

\end{center}

\begin{abstract}

Multidimensional cosmological-type  model
with  $n$  Einstein factor spaces in the theory with $l$ scalar fields and
multiple  exponential potential
is considered.  The dynamics of the model near the singularity is reduced
to a billiard on the $(N-1)$-dimensional Lobachevsky space $H^{N-1}$,
$N = n+l$. It is shown that for $n > 1$ the
oscillating behaviour near the singularity is absent
and solutions have an asymptotical Kasner-like behavior.
For the case of one scale factor ($n =1$)
billiards with finite volumes
(e.g. coinciding with that of the Bianchi-IX model)
are described and oscillating behaviour of scalar fields
near the singularity is obtained.

\end{abstract}

\bigskip

\hspace*{0.950cm} PACS number(s):\ 04.50.+h,\ 98.80.Hw,\

\np

\section{Introduction}
\setcounter{equation}{0}

The study of different aspects of multidimensional models in
gravitation and cosmology in arbitrary dimensions and with sources
as fluids and fields we started more than a decade ago
(see \cite{M0,M2,M3}). Special attention was devoted to the treatment
of dilatonic interactions with electromagnetic fields and fields of
forms of arbitrary rank \cite{IMrev}.
Here we continue our investigations of
multidimensional models, in particular  with
multiple  exponential potential (MEP) \cite{IMS}
(for $D = 4$ case see \cite{DGM}).

The  models of such sort  are currently rather popular (see, for
example, \cite{DGM,RP,CopLW,BCopN,Town} and refs. therein).
Such potentials arise naturally in certain supergravitational
models \cite{Town},   in sigma-models \cite{IMC} related to
configurations with $p$-branes and in reconstruction from observations
schemes \cite{Star}. They also appear when certain
$f(R)$ generalizations of Einstein-Hilbert action are considered
\cite{NO}.

Like in \cite{IMS}, here we consider $D$-dimensional model
governed by the action

\bear{2.1i}
    S_{act} =  \int_{M} d^{D}Z \sqrt{|g|} \{
    R[g] - h_{\alpha\beta} g^{MN} \partial_{M}\varphi^\alpha\partial_{N}
    \varphi^\beta - 2 V_{\varphi}(\varphi) \} + S_{GH},
\ear
 $D > 2$, with the scalar potential (MEP)

\beq{pot}
    V_{\varphi}(\varphi) =
    \sum_{s \in S} \Lambda_s \exp[ 2 \lambda_s (\varphi) ] .
\eeq

$S_{\rm GH}$ is the standard Gibbons-Hawking boundary term
\cite{GH}.

\begin{list}{*}{The notations used are the following ones:}
\item
    $\varphi=(\varphi^\alpha)$ is
    the vector from scalar fields in the space $\R^l$ with a metric
    determined by a non-degenerate $l\times l$ matrix
    $(h_{\alpha\beta})$ with the inverse one $(h^{\alpha\beta})$;
    $\alpha, \beta =1,\ldots,l$;
\item
     $\Lambda_s$ are constant terms; $s \in S$;
\item
    $\lambda_s$ is an $1$-form on $\R^l$:
    $\lambda_s (\varphi) =\lambda_{s \alpha} \varphi^\alpha$;\
    $\lambda^{\alpha}_{s} = h^{\alpha \beta} \lambda_{s \beta }$;
\item
    $g = g_{MN} dZ^{M} \otimes dZ^{N}$ is a metric,
    $|g| = |\det(g_{MN})|$, \\
    $M, N$  are world indices
    that may be numerated by $1, \ldots, D$.

\item
    $i, j = 1, \ldots, n$ are indices describing
    a chain of factor spaces;\\
    $A = i, \alpha$ and  $B = j, \beta$ are minisuperspace
    indices, \\
    that may be numerated also by $1, \ldots, n+l$.

\end{list}

This paper is devoted to the investigation of the possible
oscillating (and probably stochastic) behaviour near the
singularity (see \cite{BLK}-\cite{DHN} and references therein)
for cosmological type solutions corresponding to the action
(\ref{2.1i}).

We remind that near the singularity one can have an oscillating
behavior like in the well-known mixmaster  (Bianchi-IX) model
\cite{BLK}-\cite{MTW} (see also \cite{K1}-\cite{M1}).
Multidimensional generalizations and analogues
 of this model were considered by many authors (see, for
example, \cite{BK}-\cite{CDDQ}). In \cite{IKM1,IKM,IM} a billiard
representation for multidimensional cosmological models near the
singularity was considered and the criterion for a volume of the
billiard to be finite was established in terms of illumination of the unit
sphere by point-like sources. For multicomponent
perfect-fluid this was considered in
detail in \cite{IM} and generalized to $p$-brane case in \cite{IMb}
(see also \cite{DHN} and refs. therein).
Some topics related to general
(non-homogeneous) situation were considered in \cite{KM1,KM}.

Here we apply the billiard approach suggested in
\cite{IKM1,IKM,IM} to a cosmological model with MEP. We show
that (as for the exact solutions from \cite{IMS}) for $n > 1$ the
oscillating behaviour near the singularity is absent. For $n = 1$
we find here examples of oscillating behavior for scalar fields
but not for a scale factor.

The paper is organized as follows. In Sec. 2 the cosmological
model with MEP is considered:
Lagrange representation to equations of motion
and the diagonalization of the Lagrangian are presented.
In Sec. 3 a billiard approach in the
multidimensional cosmology with MEP is obtained,
and the case  $n > 1$ is studied.
Sec. 4 is devoted to description of billiards
with finite volumes in the case of one scale factor
($n = 1$).

\section{\bf The model}

Let

  \beq{2.10g}
      M = \R  \times M_{1} \times \ldots \times M_{n}
  \eeq
be a manifold equipped  with the metric

  \beq{2.11g}
     g= w \e^{2{\gamma}(u)} du \otimes du +
     \sum_{i=1}^{n} \e^{2\phi^i(u)} g^i ,
  \eeq
     where $w=\pm 1$, $u$ is a distinguished coordinate;
     $g^i$ is a  metric on
     $d_{i}$-dimensional manifold $M_i$, obeying:

  \beq{2.12g}
     {\rm Ric}[g^i ] = \xi_{i} g^i,
  \eeq
 $\xi_i= \const$,
 $i=1,\dots,n$. Thus, $(M_i,g^i)$ are Einstein spaces.

For dilatonic scalar fields we put
  \beq{2.30n}
     \varphi^\alpha=\varphi^\alpha(u),
  \eeq

\subsection{\bf Lagrangian representation}

It may be verified that the equations of motion
(see Appendix A) corresponding to
(\ref{2.1i}) for the field configuration
(\ref{2.11g})-(\ref{2.30n}) are equivalent to equations of motion for
1-dimensional $\sigma$-model with the action

 \beq{2.25gn}
     S_{\sigma} = \frac{1}2
     \int du {\cal N} \biggl\{G_{ij}\dot\phi^i\dot\phi^j
     + h_{\alpha\beta}\dot\varphi^{\alpha}\dot\varphi^{\beta}
    -2{\cal N}^{-2}V \biggr\},
 \eeq
     where $\dot x\equiv dx/du$,
 \beq{2.27gn}
     V =  -w V_{\varphi}(\varphi) \e^{2\gamma_0(\phi)}
     +  \frac w2\sum_{i =1}^{n} \xi_i d_i
        \e^{-2 \phi^i + 2 {\gamma_0}(\phi)}
  \eeq
     is the potential ($V_{\varphi}$ is defined in (\ref{pot})) with
 \beq{2.24gn}
     \gamma_0(\phi)
    \equiv \sum_{i=1}^{n}d_i\phi^i,
 \eeq
     and
 \beq{2.24gn1}
     {\cal N}=\exp(\gamma_0-\gamma)>0
 \eeq
     is the lapse function. Here
 \beq{2.c}
     G_{ij}=d_i\delta_{ij}-d_id_j, \qquad
    G^{ij}=\frac{\delta^{ij}}{d_i}+\frac1{2-D},
  \eeq
    $i,j=1,\dots,n$,
 are components of  a gravitational part of minisupermetric and its dual
 \cite{IMZ}.

\subsection{\bf Minisuperspace notations}

In what follows we consider minisuperspace $\R^{n+l}$ with points
 \beq{2.10}
     x \equiv(x^A)=(\phi^i,\varphi^\alpha)
 \eeq
     equipped by  minisuperspace metric $\bar G = \bar G_{AB}dx^A\otimes
   dx^B$ defined by the matrix and inverse one as follows:

 \bear{2.35n}
     (\bar G_{AB})=\left(\begin{array}{cc}
     G_{ij}&0\\
     0&h_{\alpha\beta}
     \end{array}\right),\quad
     (\bar G^{AB})=\left(\begin{array}{cc}
     G^{ij}&0\\
     0&h^{\alpha\beta}
     \end{array}\right).
 \ear

The potential (\ref{2.27gn}) reads

 \beq{2.40n}
    V= - w  \sum_{s \in S} \Lambda_s\e^{2U^s(x)}
    + \sum_{j=1}^n\frac w2\xi_jd_j \e^{2U^j(x)},
 \eeq
    where $U^s(x)=U_A^sx^A$ and $U^j(x)=U_A^jx^A$ are defined as
 \bear{2.u}
     U^s(x)= \lambda_{s\alpha}\varphi^{\alpha} + \gamma_0(\phi),
     \\
     \label{2.uu}
     U^j(x)=-\phi^j+\gamma_0(\phi),
  \ear
    or, in components,
  \bear{2.38n}
    (U_A^s)=(d_i,\lambda_{s \alpha})
    \\  \label{2.43n}
    (U_A^j)=(-\delta_i^j+d_i,0)
  \ear
  $s \in S$; $i,j=1,\dots,n$.

The integrability of the Lagrange system (\ref{2.25gn})
depends upon the scalar products of co-vectors $U^s$, $U^i$ corresponding
to $\bar G$:

\beq{2.45n}
    (U,U')=\bar G^{AB}U_AU'_B,
\eeq
    These products have the following form
\bear{2.48n}
   (U^i,U^j)=\frac{\delta_{ij}}{d_j}-1,
    \\ \label{2.51n}
    (U^s,U^{s'})= - b + \lambda_{s}\cdot\lambda_{s'},
    \\ \label{2.52n}
    (U^s,U^i)= -1,
\ear
    where
\bear{2.55n}
    \lambda_s\cdot\lambda_{s'} \equiv
    \lambda_{s\alpha}\lambda_{s' \beta}h^{\alpha\beta},
    \quad     b = \frac{D-1}{D-2},
\ear
    $s, s' \in S$.

\subsection{ Diagonalization of the Lagrangian}

Let the matrix $(h_{\alpha \beta})$ have the Euclidean signature.
Then, the  minisuperspace metric (\ref{2.35n})
has a pseudo-Euclidean signature \\
   $(-,+, \ldots ,+)$  since the matrix $(G_{ij})$ has the
pseudo-Euclidean signature \cite{IMZ}.

Hence there exists
a linear transformation
 \beq{2.21o}
 z^{a}=e^{a}_{A}x^{A},
 \eeq
diagonalizing the minisuperspace metric (\ref{2.35n})
 \beq{2.22o}
 \bar G= \eta_{ac}dz^{a} \otimes dz^{c}=
  -dz^{0} \otimes dz^{0} + \sum_{k=1}^{N-1}dz^{k}\otimes dz^{k},
 \eeq
where
 \beq{2.23o}
 (\eta_{ac})=(\eta^{ac}) \equiv {\rm diag}(-1,+1, \ldots ,+1),
 \eeq
and here and in what follows
 $a,c = 0, \ldots ,N-1$; $N =n+l$.
The matrix of  linear transformation
 $(e^{a}_{A})$  satisfies the relation
 \beq{2.24o}
 \eta_{ac} e^{a}_{A} e^{c}_{B} = \bar{G}_{AB}
 \eeq
or, equivalently,
  \beq{2.25o}
  \eta^{ac} = e^{a}_{A}\bar{G}^{AB} e^{c}_{B} =  (e^{a},e^{c}),
  \eeq
where $e^a = (e^a_A)$.

Inverting the map (\ref{2.21o}) we get
  \beq{2.28o}
  x^{A} = e_{a}^{A} z^{a},
  \eeq
where for components of the inverse matrix
 $(e_{a}^{A}) = (e^{a}_{A})^{-1}$ we obtain from (\ref{2.25o})
  \beq{2.29o}
  e_{a}^{A}    = \bar{G}^{AB} e^{c}_{B} \eta_{ca}.
  \eeq

Like in \cite{IM} we put
  \beq{2.30o}
  e^{0} = q^{-1} U^{\Lambda}, \qquad
  q = [- (U^\Lambda,U^\Lambda)]^{1/2} = b^{1/2}.
  \eeq
where $U^\Lambda(x)=U_A^\Lambda x^A=\gamma_0(\phi)$ is
co-vector corresponding to the cosmological term,
or, in components
    \bear{2.44n}
    (U_A^\Lambda)=(d_i,0),
    \ear

and hence
  \beq{2.31o}
  z^0 = e^{0}_{A} x^A = \sum_{i=1}^{n} q^{-1} d_i x^i.
  \eeq

In $z$-coordinates (\ref{2.21o}) with $z^0$ from
(\ref{2.31o}) the Lagrangian
  corresponding to (\ref{2.25gn}) reads
  \beq{2.32o}
  L = {L}(z, \dot{z}, {\cal N})
  = \frac{1}{2} {\cal N}^{-1}
    \eta_{ac} \dot{z}^{a} \dot{z}^{c} -  {\cal N} V(z),
  \eeq

where
  \beq{2.34o}
  {V}(z) =
  \sum_{r \in S_{*}} A_{r} \exp(2 u^{r}_a z^a)
  \eeq
is a potential,
  \beq{2.34oa}
  S_{*} = \{ 1, \ldots, n \} \cup S
  \eeq
is an extended index set and
  \beq{2.34ob}
  A_j = \frac{w}{2} \xi_j d_j, \quad
  A_s = -w \Lambda_s,
\eeq
  $j = 1, \ldots, n$; $s \in S$.
Here we denote
  \beq{2.35o}
  u^{r}_a  = e_{a}^{A} U^{r}_A =
  (U^{r}, e^{c}) \eta_{ca},
\eeq
  $a = 0, \ldots , N-1$; $r \in S_{*}$ (see (\ref{2.29o})).

From (\ref{2.45n}), (\ref{2.30o}) and (\ref{2.35o})
we deduce
  \beq{2.36o}
  u^{r}_0  = - (U^{r}, e^{0}) =
  (\sum_{i=1}^{n} U^{r}_i ) / q(D-2),
  \eeq
 $r \in S_*$.

For the potential-term and curvature-term components
we obtain from  (\ref{2.30o})  and (\ref{2.36o})
  \beq{2.37o}
    u^{s}_0 = q > 0 , \qquad u^{j}_0 = 1/q > 0,
  \eeq
 $j= 1,\ldots,n$.

We remind that  (see (\ref{2.48n}))
  \beq{2.39o}
  (U^{j},U^{j}) = \left(\frac{1}{d_j} - 1 \right) < 0,
  \eeq
for $d_j > 1$, $j= 1,\ldots,n$. For $d_j = 1$ we have
    $\xi^{j} = A_{j} = 0$.

\section{Billiard representation}
\setcounter{equation}{0}

Here we put the following restriction on parameters of the model:
  \bear{4.1n}
   -w \Lambda_s > 0,
   \\ \label{4.2n}
   {\rm if} \quad
   (U^s,U^{s})= -b  + \lambda_s^2 >0,
  \ear
    $s \in S$.  In what follows we denote by $S_{+}$
a subset of all $s \in S$ satisfying (\ref{4.2n}).
As we shall see below these restrictions are necessary
for a formation of billiard ``walls''
(with positive infinite potential) in approaching to singularity.

Due to relations
 (\ref{2.34ob}), (\ref{2.37o}), (\ref{2.39o})
and  (\ref{4.1n})  the parameters $u^r_a$  in the potential (\ref{2.34o})
obey the following restrictions:
  \bear{3.3o}
  && {\bf 1.} \ A_r > 0  \
  {\rm for} \ (u^r)^2 = -(u^r_0)^2 + (\vec{u}^r)^2 > 0;
   \\  \label{3.4o}
  && {\bf 2.} \ u^r_0 > 0  \ {\rm for \ all} \  r \in S_{*}.
  \ear

Due to relations  (\ref{3.3o}) and (\ref{3.4o}) the Lagrange
system (\ref{2.32o}) for $N \geq 3$ in the ("near the
singularity") limit
  \beq{3.1o}
   z^0  \rightarrow  -\infty, \qquad z^0 < - |\vec{z}|,
  \eeq
may be reduced to a motion of a point-like
particle in $N-1$-dimensional billiard
belonging to Lobachevsky space \cite{IKM1,IKM,IM,IMb}.

For non-exceptional asymptotics (non-Milne-type)
the limit (\ref{3.1o})
describes the approaching to the singularity.
(in this case  the volume scale factor vanishes
   $\exp(\sum_{i=1}^{n} d_{i}x^{i}) = \exp(qz^0)
   \rightarrow + 0$).

Indeed, introducing generalized  Misner-Chitre coordinates
in the lower light cone $z^0 < - |\vec{z}|$ \cite{IKM1,IKM}

  \bear{3.6o}
  &&z^0 = - \exp(-y^0) \frac{1 + \vec{y}^2}{1 - \vec{y}^2}, \\
  &&\vec{z} = - 2 \exp(-y^0) \frac{ \vec{y}}{1 - \vec{y}^2},
  \ear
  $|\vec{y}| < 1$, and fixing the time gauge

  \beq{3.13o}
  {\cal N} =   \exp(- 2y^0) = - z^2.
  \eeq
  we get in the limit $y^{0} \rightarrow - \infty$
 (after separating $y^0$ variable)

  a "billiard"  Lagrangian
  \beq{3.33o}
  L_{B} =  \frac{1}{2} \bar{h}_{ij}(\vec{y})
  \dot{y}^{i} \dot{y}^{j} -  V(\vec{y},B).
  \eeq

Here
  \beq{3.9o}
  \bar{h}_{ij}(\vec{y}) = 4 \delta_{ij} (1 - \vec{y}^2)^{-2},
  \eeq
  $i,j =1, \ldots , N-1$, are components of the canonical metric on
 the $(N-1)$-dimensional Lobachevsky space
 $H^{N-1}= D^{N-1} \equiv \{ \vec{y}: |\vec{y}| < 1 \}$.

The "wall" potential $V(\vec{y},B)$ in
(\ref{3.33o})
  \bear{3.25o}
   V(\vec{y},B) \equiv &0, &\vec{y} \in B,
  \nonumber \\
  &+ \infty, &\vec{y}
  \in D^{N-1} \setminus B,
  \ear
corresponds to the open domain (billiard)
  \beq{4.4}
   B = \bigcap_{s \in S_{+}} {B}_{s}  \subset D^{N-1},
   \eeq
where
  \beq{4.5}
  {B}_{s}  = \{ \vec{y} \in D^{N-1} |
  |\vec{y} - \vec{v}^{s}| > r_{s} \},
  \eeq
and
   \beq{4.6}
   \vec{v}_s = - \vec{u}^s/u^s_0, \qquad
   r_s = \sqrt{(\vec{v}_s)^2 - 1},
   \eeq
 ($|\vec{v_s}| > 1$)  $s \in S_{+}$.

The boundary of the billiard is formed by certain parts of $m_{+}
 = |S_{+}|$ $(N-2)$-dimensional spheres with centers in points
 $\vec{v}_s$ and radii $r_s$, $s \in S_{+}$.

When  $S_{+} \neq \emptyset$ the Lagrangian
 (\ref{3.33o}) describes a motion of a particle  of  unit  mass,
moving in the ($N-1$)-dimensional billiard $B \subset D^{N-1}$  (see
 (\ref{4.4})).  The geodesic motion in $B$
corresponds to a "Kasner epoch" while the reflection from the boundary
corresponds to the change of Kasner epochs.

The billiard $B$  has an infinite volume: ${\rm vol} B = +\infty$
if and only if there are open zones at the infinite sphere
  $|\vec{y}|  = 1$. After a finite number of reflections from the boundary
a  particle  moves  towards  one  of  these  open  zones.
In this case for a
corresponding cosmological model we get the "Kasner-like"
behavior in the limit $t \rightarrow - \infty$ \cite{IMb}.

When ${\rm vol} B < + \infty$ we get
a never ending oscillating behaviour near the singularity.

In \cite{IM} the following simple geometric criterion  for
the finiteness of  volume of $B$ was proposed.

 {\bf Proposition 1 \cite{IM}.}
 {\em The billiard $B$ (\ref{4.4}) has a finite
volume if and only if  point-like
sources of light located at the points $\vec{v}_s$
 $s \in S_{+}$ (see (\ref{4.6}))
illuminate the unit sphere $S^{N-2}$}.

There exists a topological bound on a number of point-like
sources $m_{+}$  illuminating the sphere $S^{N-2}$
 \cite{BG}:
 \beq{3.40o}
 m_{+} \geq N.
 \eeq

Due to this restriction the number of
exponential terms in potential obeying (\ref{4.2n})
  $m_{+} =|S_{+}|$
should at least exceed the value $N = n+l$ for
the existence of oscillating  behaviour
near the singularity.

 {\bf Description in terms of Kasner-like parameters.}

For zero potential $V_{\varphi} = 0$ we get a Kasner-like
solutions
 \bear{4.8}
   &&g = w d\tau \otimes d\tau + \sum_{i=1}^{n} A_i \tau^{2 \alpha^i}
   g^i,
   \\  \label{4.9}
   &&\varphi^{\beta} =  \alpha^{\beta} \ln \tau +
   \varphi^{\beta}_0,
   \\  \label{4.10}
   && \sum_{i=1}^{n} d_i \alpha^i =
   \sum_{i=1}^{n} d_i (\alpha^i)^2 +
  \alpha^{\beta} \alpha^{\gamma} h_{\beta \gamma}= 1,
 \ear
where $A_i > 0$ and  $\varphi^{\beta}_0$ are constants,
 $i = 1, \ldots, n$; $\beta, \gamma = 1, \ldots, l$.

Let $\alpha = (\alpha^{A}) =
(\alpha^{i}, \alpha^{\gamma})$ obey the relations
 \beq{4.12}
     U^s(\alpha) =  U_A^{s} \alpha^A = \sum_{i =1}^{n} d_i\alpha^i
     + \lambda_{a_s \gamma}\alpha^{\gamma} > 0,
   \eeq
for all $s \in S_{+}$,
then the field configuration (\ref{4.8})-(\ref{4.10})
is the  asymptotical (attractor) solution for a family of (exact)
solutions,  when $\tau \to +0$.

Relations (\ref{4.12})  may be  easily
understood  using the following relations
  \beq{4.12b}
  \Lambda_s \exp[ 2 \lambda_s (\varphi) + 2 \gamma_0(\phi)] =
  \Lambda_s \exp[2U^s(x)] = C_s \tau^{2U^s(\alpha)} \to 0,
    \eeq
for $\tau \to +0$, where $C_s \neq 0$ are constants, $s \in
 S_{+}$. Other terms in the potential (\ref{2.27gn}) are also
vanishing near the singularity \cite{IKM1,IKM,IM,IMb}. Thus, the
potential (\ref{2.27gn}) asymptotically tends to zero as $\tau \to
+0$ and we are led to asymptotical solutions
(\ref{4.8})-(\ref{4.10}).

Another way to get the conditions (\ref{4.12})
is based on the isomorphism between
$S^{N-2}$ and the Kasner set  (\ref{4.10})
  \beq{4.13}
  \alpha^A =  e^A_a n^a / q, \quad  (n^a) = (1, \vec{n}),
  \quad \vec{n} \in S^{N-2}.
  \eeq
Here we use the diagonalizing matrix $(e^A_a)$ and the parameter
 $q$ defined in the previous section (see (\ref{2.30o}))
 \cite{IM,IMb}. Thus, we come to the following proposition.

 {\bf Proposition 2.}
 {\em Billiard $B$ (\ref{4.4}) has a finite volume if and only if
there are no $\alpha$  satisfying the
relations  (\ref{4.10}) and (\ref{4.12}).}

So, we obtained a billiard representation for the model under
consideration when the restrictions (\ref{4.1n})  are  imposed.

Here we present also useful relations  describing  the billiard in
terms of scalar products
    \bear{4.8b}
    \vec{v}_{s} \vec{v}_{s'} =
    \frac{\vec{u}^{s} \vec{u}^{s'}}{ u^{s}_0 u^{s'}_0}
     = b^{-1}  \lambda_{s \alpha} \lambda_{s' \beta}
    h^{\alpha\beta},
  \ear
 $s, s' \in S_{+}$. They follow from the formulas
 $\vec{u}^{s} \vec{u}^{s'} - u^{s}_0 u^{s'}_0 = (U^s,U^{s'})$
and (\ref{2.51n}).

{\bf Proposition 3.}
{\em For $n > 1$ billiard $B$ (\ref{4.4}) has an  infinite volume.}

{\bf Proof. } Due to Proposition 2 it is sufficient to present at
least one set of Kasner parameters $\alpha = (\alpha^{i},
\alpha^{\gamma})$  obeying the relations (\ref{4.10}) and
(\ref{4.12}). As an example of such set one may choose any Kasner
set $\alpha$ (obeying   (\ref{4.10}) ) with $\alpha^{\gamma} = 0$,
for example, with the following components
  \beq{4.8c}
  \alpha^1 = \frac{d_1 \pm \sqrt{R}}{d_1 (d_1 + d_2)}, \quad
  \alpha^2 = \frac{d_2 \mp  \sqrt{R}}{d_2 (d_1 + d_2)}, \quad
  \alpha^i = 0 \quad (i > 2 ).
  \eeq
where $R = d_1 d_2 (d_1 + d_2 -1)$.
In this case inequalities (\ref{4.12}) are satisfied,
since $ U^s(\alpha) = 1$ for all $s$.
The proposition is proved.

Thus, according to Proposition 3,
for  $n > 1$  we obviously have a "Kasner-like"
behavior near the singularity  (as  $\tau \rightarrow 0$).
The oscillating behaviour near the singularity is impossible
in this case.

   {\bf Remark 1 (general ``collision law'').}
The set of Kasner parameters  $(\alpha^{'A})$
after the collision with the $s$-th wall ($s \in S_{+}$) is defined
by the Kasner set before the collision $(\alpha^{A})$ according
to the following formula  \cite{Ierice}
  \beq{gcl}
  \alpha^{'A} =
               \frac{\alpha^A - 2 U^s(\alpha) U^{sA}(U^s,U^s)^{-1}}
               {1 - 2 U^s(\alpha) (U^s,U^{\Lambda})(U^s,U^s)^{-1}},
  \eeq
where $U^{s A} = \bar{G}^{AB} U^s_B$,
 $U^s(\alpha) =  U_A^{s} \alpha^A$ and co-vector $U^{\Lambda}$
is defined in (\ref{2.44n}). In the special case of one
scalar field and  $1$-dimensional factor-spaces
(i.e.  $l= d_i =1$) this formula was suggested earlier
in \cite{DH}.

\section{One factor-space}

In this section we consider examples of $l$-dimensional
billiards with finite volumes
that occur in the model with  $l$-scalar fields ($l \geq 2$)
and one scale factor ($n= 1$). Here we put
 $h_{\alpha\beta} = \delta_{\alpha \beta}$
and $\vec{\lambda}_s = (\lambda_{s1}, \dots, \lambda_{sl})$.

Thus, here we deal with the Lagrangian
 \beq{5.1i}
    {\cal L} =
    R[g] -  \partial_{M}\vec{\varphi} \partial_{N} \vec{\varphi} -
    2 \sum_{s \in S} \Lambda_s  \exp[ 2 \vec{\lambda}\vec{\varphi}].
 \eeq
where  $\vec{\varphi} = (\varphi^1, \dots, \varphi^l)$.

In this (one-factor case) the following proposition
takes place.

 {\bf Proposition 4.}
 {\em For $n =1$ and
 $h_{\alpha\beta} = \delta_{\alpha \beta}$
 the billiard $B$ (\ref{4.4}) has a finite
volume if and only if  point-like
sources of light located at the points
   $b^{-1/2} \vec{\lambda}_s \in \R^l$,
   $s \in S_{+}$, ($b = (D-1)/(D-2)$) illuminate the unit sphere
   $S^{l-1}$ }.

{\bf Proof.}
According to relations (\ref{4.8b}) the
set of vectors
 $b^{-1/2} \vec{\lambda}_s \in \R^l$, $s \in S_{+}$,
is coinciding with the set
 $\vec{v}_s \in \R^l$, $s \in S_{+}$, up to
 $O(l)$-transformation, i.e. there exists
orthogonal matrix $A$:  $A^{T} A = {\bf 1}$,
such that
   $b^{-1/2} \vec{\lambda}_s = A \vec{v}_s \in \R^l$, $s \in S_{+}$.
   Then the Proposition 4 follows from Proposition 1,
   since the  sphere $S^{l-1}$ is illuminated
   by sources $\vec{v}_s$, $s \in S_{+}$,
   if and only if, it is illuminated by
   sources $b^{-1/2} \vec{\lambda}_s$, $s \in S_{+}$.

 According to  relations (\ref{4.8})-(\ref{4.10})
 we get the following
 asymptotical behavior for $\tau \to 0$

 \bear{5.8}
   &&g_{as} = w d\tau \otimes d\tau +  A_1 \tau^{2/(D-1)} g^1,
   \\  \label{5.9}
   &&\vec{\varphi}_{as} =  \vec{\alpha}_{\varphi} \ln \tau +
   \vec{\varphi}_0,
   \\  \label{5.10}
   &&(\vec{\alpha}_{\varphi})^2 = b^{-1} = (D-2)/(D-1).
 \ear

 Here   $\vec{\varphi}_0$ and $\vec{\alpha}_{\varphi}$ change
 their values after the reflections from the billiard
 walls. Thus, here we obtained the oscillating
 behaviour of scalar fields near the singularity.

   {\bf Remark 2 (``collision law'').}
From (\ref{gcl}) we get the ``collision law''
relation in this case
  \beq{cl}
   \vec{\alpha}^{'}_{\varphi} =
   \frac{\vec{\alpha}_{\varphi} - 2 (1 + \vec{\lambda}_s
   \vec{\alpha}_{\varphi})(\lambda_s^2 - b)^{-1} \vec{\lambda}_s}
   {1 +  2 (1 + \vec{\lambda}_s
   \vec{\alpha}_{\varphi})(\lambda_s^2 - b)^{-1} b}.
  \eeq
The Kasner parameter for the scale factor is not changed
after the ``collision''.

 {\bf $l =2$ case.} In the special case
of two-component scalar field
 ($l =2)$, and   $m_{+} =|S_{+}| = 3$
(i.e. when three ``walls'' appear)
 we find the necessary and sufficient
 condition for the finiteness of the billiard
 volume in terms of scalar products of
 the coupling vectors
 $\vec{\lambda}_s \in \R^2$, $s \in S_{+}$.

 {\bf Proposition 5 .}
 {\em For $n =1$, $l =2$,
 $h_{\alpha\beta} = \delta_{\alpha \beta}$
 and   $m_{+} =|S_{+}| = 3$
 the billiard $B$ (\ref{4.4}) has a finite
 volume if and only if  the vectors
 $\vec{\lambda}_s \in \R^2$, $s \in S_{+}$,
 obey the following  relations:

 \bear{5.11}
 b^{-1} \vec{\lambda}_s  \vec{\lambda}_{s'}
 \geq  1 - \sqrt{b^{-1} \vec{\lambda}_s^2 -1}
 \sqrt{b^{-1} \vec{\lambda}_{s'}^2 -1},
  \qquad s < s',
 \\   \label{5.12}
  \sum_{s < s'} \arccos
  \frac{\vec{\lambda}_s \vec{\lambda}_{s'}}
  {|\vec{\lambda}_s| |\vec{\lambda}_{s'}|}  = 2 \pi.
  \ear
 }

{\bf Proof.}
According to Proposition 3 we should find
the necessary and sufficient conditions
for three points located in
 $\vec{v}_s = b^{-1/2} \vec{\lambda}_s$, $s = 1,2,3$,
to illuminate the unit circle $S^1$.
Here we put $S_{+} = \{1,2,3 \}$ for simplicity.
It may
be obtained from a simple geometrical
consideration that such conditions may
be chosen as the following ones:

  \beq{5.13}
   \theta_{s s'} \leq  \arccos \frac{1}{|\vec{v_s}|}
                  +  \arccos \frac{1}{|\vec{v_{s'}}|},
   \qquad s < s',
   \eeq
and
    \beq{5.14}
   \theta_{12} + \theta_{23} + \theta_{13} = 2 \pi
    \eeq
where
   \beq{5.15}
   \theta_{s s'} =
    \arccos \frac{\vec{v}_s \vec{v}_{s'} }{|\vec{v}_s|| \vec{v}_{s'}|}
    \eeq
is the angle between vectors  $\vec{v_s}$ and $ \vec{v_{s'}}$
 $s,s' =1,2,3$. Relation (\ref{5.13}) means that the angle
between two vectors  $\vec{v_s}$ and $ \vec{v_{s'}}$ should not
exceed one half of the sum of two arcs on $S^1$ illuminated
by source of light  located in points  $\vec{v_s}$ and $ \vec{v_{s'}}$
(see Fig. 1).

Relations  (\ref{5.11}) may be obtained from
(\ref{5.13}) by acting on both sides of the inequality
by function $cos$. Relation (\ref{5.12}) is just equivalent
to (\ref{5.13}). Relation (\ref{5.14}) exclude
the situation when points $ \vec{v_s}$,  $s =1,2,3$,
belong to a half-plane with a border-line containing
the center of the unit circle.
The proposition is proved.

%
%

\Picture{82}{71}{pic2.gif}{%
Triangle sub-compact billiard with finite volume for
$n= 1$, $l =2$ and  $m_{+} = 3$ .}

An example of (sub-)compact triangle billiard with a finite area
in the Lobachevsky space $H^2$ is depicted on Fig. 1.

In the symmetric case when all
$\lambda_s^2 = 4b$ and
$\vec{\lambda}_s  \vec{\lambda}_{s'} = -2b$
for $s \neq s'$ we get an
example of non-(sub)-compact billiard
with finite area. Such
billiard appears in the well-known Bianchi-IX model, see Fig. 2.

For ``quasi-Cartan'' matrix defined
as $A_{ss'} = 2 (U^s,U^{s'})/(U^{s'},U^{s'})$
we get
 \beq{B.1}
  \left(A_{ss'}\right)=\left(\begin{array}{ccc}
   2 & -2 & -2 \\
  -2 &  2 & -2 \\
  -2 & -2 &  2
 \end{array}\right).
 \eeq

This matrix coincides with the Cartan matrix
of the  hyperbolic of Kac-Moody  algebra
that is number 7 in classification of \cite{Sac}.
This  Kac-Moody algebra is a subalgebra of $A_1^{\wedge \wedge}$
\cite{FN} (see also \cite{DHJN,dBS}).


\Picture{82}{71}{pic3.gif}{%
Triangle billiard  coinciding with that of Bianchi-IX
model.}


\section{Discussions}

In this paper we have considered the behavior near the singularity
of  the multidimensional cosmological-type model
with   $n$  Einstein factor-spaces in the theory with  scalar
fields and MEP (multiple  exponential potential). Using the
results from \cite{IKM1,IKM,IM,IMb} we have obtained the billiard
representation on  multidimensional Lobachevsky space for this
cosmological-type model near the singularity.

Here we have  shown that for $n > 1$ the oscillating behavior near
the singularity is absent, i.e. solutions have an asymptotical
Kasner-like behavior.

For one-factor case we have described (in terms of illumination
problem) the billiards with finite volume and hence with the
oscillating behavior of scalar fields near the singularity.

In the model with two scalar fields and  three  potential walls
we have found the necessary and sufficient conditions
(in terms of  dilatonic coupling vectors) for
triangle billiards to be of finite volume.

\renewcommand{\theequation}{\Alph{subsection}.\arabic{equation}}
\renewcommand{\thesection}{}
\renewcommand{\thesubsection}{\Alph{subsection}}
\setcounter{section}{0}

 \section{Appendix}

 \subsection{Equations of motion}

Here we outline for the sake of completeness the equations of motions
corresponding to the action (\ref{2.1i})
\bear{A.1}
    R_{MN} - \frac{1}{2} g_{MN} R  =   T_{MN} ,
    \\    \label{A.2}
    {\btu}[g] \varphi^\alpha -
    \sum_{s \in S} 2 \lambda^{\alpha}_s
    e^{2 \lambda_s(\varphi)} \Lambda_s = 0.
\ear

In (\ref{A.1})
\bear{A.3}
  T_{MN} =
  h_{\alpha\beta}\left(\p_{M} \varphi^\alpha \p_{N} \varphi^\beta -
  \frac{1}{2} g_{MN} \p_{P} \varphi^\alpha \p^{P} \varphi^\beta\right)
  - V_{\varphi} g_{MN}.
\ear

\bigskip

{\bf Acknowlegments}

This work was supported in part by the Russian Ministry of
Science and Technology, Russian Foundation for Basic Research
(RFFI-01-02-17312-a)and DFG Project (436 RUS 113/678/0-1(R)).

Authors thank colleagues from the Department of Physics,
University of Konstanz, for the
hospitality during their visits to Konstanz
in August-December, 2003.

\end{document}